\documentclass{article}

\usepackage[dvipdfmx]{graphicx}
\usepackage{amsmath}        
\usepackage{amssymb}        
\usepackage{cite}
\usepackage[dvips]{color}
\usepackage{fancyhdr}

\def\erase#1{{}}
\def\EqArrerase#1{{}}

\makeatletter
 \renewcommand{\theequation}{%
 \thesection.\arabic{equation}}
 \@addtoreset{equation}{section}
\makeatother

\def\GL{{G\kern-.12em L\kern-.04em}}
\def\OSp{{O\kern-.11em S\kern-.04em p}}
\def\IOSp{{I\kern-.06em O\kern-.11em S\kern-.04em p}}
\def\MN{{M\kern-.14em N}}
\def\NM{{N\kern-.14em M}}
\def\NL{{N\kern-.14em L}}
\def\LN{{L\kern-.11em N}}
\def\ML{{M\kern-.14em L}}
\def\LM{{L\kern-.11em M}}
\def\RN{{R\kern-.11em N}}
\def\NR{{N\kern-.14em R}}
\def\RM{{R\kern-.11em M}}
\def\MR{{M\kern-.14em R}}
\def\RL{{R\kern-.11em L}}
\def\LR{{L\kern-.11em R}}
\def\RS{{R\kern-.11em S}}
\def\SR{{S\kern-.11em R}}
\def\SN{{S\kern-.11em N}}
\def\NS{{N\kern-.11em S}}
\def\SM{{S\kern-.11em M}}
\def\MS{{M\kern-.11em S}}
\def\SL{{S\kern-.11em L}}
\def\LS{{L\kern-.11em S}}
\def\sqr#1#2{{\vcenter{\hrule height.#2pt
      \hbox{\vrule width.#2pt height#1pt \kern#1pt
          \vrule width.#2pt}
      \hrule height.#2pt}}}
\def\bra0{\langle0|}
\def\ket0{|0\rangle}
\def\soeji#1_#2#3{#1_{#2}\cdots#1_{#3}}
\def\longgLRarrow{\longleftarrow\kern-3pt\relbar\kern-3pt\relbar\kern-3pt%
\longrightarrow}
\def\longLRarrow{\longleftarrow\kern-3pt\relbar\kern-3pt\longrightarrow}
\def\longLarrow{\longleftarrow\kern-3pt\relbar\kern-3pt\relbar\kern-3pt\relbar}
\def\longRarrow{\relbar\kern-3pt\relbar\kern-3pt\relbar\kern-3pt\longrightarrow}
\def\bothDer#1#2#3{%
\overset{\kern-.7em\stackrel{#1}{#2}}{\partial_{#3}}}
 
\makeatletter
 \renewcommand{\theequation}{%
 \thesection.\arabic{equation}}
 \@addtoreset{equation}{section}
\makeatother

\begin{document}
\thispagestyle{fancy}

\title{Vanishing Metric Commutation Relation 
and 
Higher-derivative De Donder Gauge in Quadratic Gravity}

\author{Ichiro Oda
\footnote{Electronic address: ioda@cs.u-ryukyu.ac.jp}
\\
{\it\small
\begin{tabular}{c}
Department of Physics, Faculty of Science, University of the 
           Ryukyus,\\
           Nishihara, Okinawa 903-0213, Japan\\      
\end{tabular}
}
}
\date{}

\maketitle

\thispagestyle{fancy}

\begin{abstract}

We show that the equal-time commutation relations (ETCRs) among the time derivatives of the metric tensor identically vanish 
in the higher-derivative de Donder gauge as well as the conventional de Donder gauge (or harmonic gauge) for general coordinate invariance 
in the manifestly covariant canonical operator formalism of quadratic gravity. These ETCRs provide us with the vanishing 
four-dimensional commutation relation, which implies that the metric tensor behaves as if it were not a quantum operator but
a classical field. In this case, the micro-causality is valid at least for the metric tensor in an obvious manner. This fact might be a 
manifestation of renormalizability of quadratic gravity in case of the canonical operator formalism.

\end{abstract}

\newpage
\pagestyle{plain}
\pagenumbering{arabic}


\section{Introduction}

In recent years, higher-derivative gravity \cite{Stelle1}, which is also called $\it{quadratic \, gravity}$, has raised
a considerable interest \cite{Luca}. This is because many of new ideas to make sense of the massive ghost existing in quadratic gravity, 
which breaks the unitarity of physical S-matrix, have been advocated \cite{Anselmi, Salvio, Strumia, Donoghue1, Donoghue2,
Kubo-Kugo, Holdom, Luca2}. It is worth stressing that 
the unitarity is one of the most important requirements in quantum field theory so if the massive ghost problem were overcome, 
for the first time we could get a renormalizable and physically consistent quantum gravity within the framework of quadratic gravity.

However, these new ideas are almost based on Feynman's path integral approach and toy models, and their plausibility as quantum field theory
seems to be obscure. Since quantum field theory is formulated in two different but mathematically equivalent formalisms, those are path 
integral formalism and canonical operator formalism \cite{Kugo-Ojima, Nakanishi, N-O-text}, such the new ideas should be described
in terms of the canonical operator formalism as well. If they were not able to be formulated in terms of the canonical operator formalism,
such the new ideas would be simply wrong or violate some fundamental postulates of quantum field theory such as causality and unitarity etc. 
Moreover, the toy models describe only some aspects of quadratic gravity so one should understand a full theory of quadratic gravity. 

With such a philosophy,  by extending previous works \cite{Kimura1, Kimura2, Kimura3}, we have recently constructed the manifestly 
covariant canonical operator formalism, which is sometimes called the BRST formalism, of quadratic gravity and briefly commented 
on the problem of the unitarity violation within the framework of the canonical formalism \cite{Oda-Can}.\footnote{See also related works
\cite{Oda-Q, Oda-W, Oda-Saake, Oda-Corfu, Oda-Ohta, Oda-Conf, Oda-f}.}
In this study, we have found a remarkable feature of quadratic gravity in the canonical formalism, which is the vanishing equal-time 
commutation relations (ETCRs) among the time derivatives of the metric tensor. It is of interest to recall that these ETCRs possess 
a nontrivial expression in the other higher-derivative gravity such as conformal gravity \cite{Oda-Ohta, Oda-Conf}
and $f(R)$ gravity \cite{Oda-f} in addition to Einstein's general relativity \cite{Nakanishi, N-O-text}. Thus, it is expected that this phenomenon 
might reflect the important peculiar feature of quadratic gravity such as renormalizability.

However, one anxiety is that the vanishing ETCRs among the time derivatives of the metric tensor might be an artifact of the choice 
of gauge conditions, that is the choice of the de Donder gauge condition for general coordinate invariance.
Moreover, since it is natural to adopt a higher-derivative gauge fixing condition in quadratic gravity \cite{Fradkin}, we should investigate 
whether such a remarkable feature of the vanishing ETCRs is still valid or not in case of the higher-derivative gauge conditions.
 
From this viewpoint, in this article we would like to adopt the higher-derivative de Donder gauge condition and show that the vanishing
ETCRs of the metric are still valid in this gauge choice as well. The reason why we limit ourselves to only the de Donder gauge is 
that the theory becomes remarkably beautiful in this specific choice of gauge-fixing. For instance, in the de Donder gauge 
all the ETCRs can be obtained explicitly in closed form. This cannot be never achieved in any preceding attempt in 
quantum gravity. Furthermore, together with the $GL(4)$ symmetry, the de Donder gauge not only gives us a very huge global symmetry 
named $IOSp(8|8)$ \cite{Nakanishi, N-O-text} and its extended symmetry \cite{Oda-W, Oda-Saake, Oda-Ohta, Oda-Conf}, 
which includes the BRST symmetry and translational symmetry etc., but also one can prove 
that the graviton must be precisely massless via spontaneous symmetry of $GL(4)$ symmetry to the $SO(1, 3) $ Lorentz 
symmetry \cite{NO-grav}.\footnote{Provided that there is a local scale symmetry, one can show that the dilaton is exactly massless along the similar 
argument \cite{Oda-W, Oda-Saake, Oda-Ohta, Oda-Conf}.}   
          
We close this section with an overview of this article. In Section 2, we review the manifestly covariant canonical operator formalism
of quadratic gravity in the de Donder gauge \cite{Oda-Can}. In Section 3, we construct the higher-derivative de Donder gauge on the basis of the BRST
formalism. In Section 4, we show that the ETCRs among time derivatives of the metric tensor vanish identically even in the higher-derivative 
de Donder gauge. In Section 5, we derive the vanishing four-dimensional commutation relation of the metric tensor from the vanishing ETCRs
of the metric tensor and consider its physical implication. The final section is devoted to the conclusion.

\section{Review of canonical operator formalism of quadratic gravity}

In this section, we start by reviewing our previous study of the canonical operator formalism of quadratic gravity \cite{Oda-Can}.
The classical Lagrangian takes the form\footnote{We follow the notation and conventions of Misner-Thorne-Wheeler (MTW) textbook \cite{MTW}. 
Lowercase Greek letters $\mu, \nu, \dots$ and Latin ones $i, j, \dots$ are used for space-time and spatial indices, respectively; 
for instance, $\mu= 0, 1, 2, 3$ and $i = 1, 2, 3$. The Riemann curvature tensor and the Ricci tensor are, respectively, defined by 
$R^\rho{}_{\sigma\mu\nu} = \partial_\mu \Gamma^\rho_{\sigma\nu} 
- \partial_\nu \Gamma^\rho_{\sigma\mu} + \Gamma^\rho_{\lambda\mu} \Gamma^\lambda_{\sigma\nu} 
- \Gamma^\rho_{\lambda\nu} \Gamma^\lambda_{\sigma\mu}$ and $R_{\mu\nu} = R^\rho{}_{\mu\rho\nu}$. 
The Minkowski metric tensor is denoted by $\eta_{\mu\nu}$; $\eta_{00} = - \eta_{11} = - \eta_{22} 
= - \eta_{33} = -1$ and $\eta_{\mu\nu} = 0$ for $\mu \neq \nu$. Moreover, $x^\mu = ( x^0, x^i ) = ( t, x^i )$
and $p^\mu = ( p^0, p^i ) = ( E, p^i )$.} 
\begin{eqnarray}
{\cal L}_0 = \sqrt{-g} \left( \frac{1}{2 \kappa^2} R + \alpha_R R^2 + \alpha_C C_{\mu\nu\rho\sigma} C^{\mu\nu\rho\sigma} \right),
\label{Class-Lag1}  
\end{eqnarray}
where $\kappa$ is defined as $\kappa^2 = 8 \pi G = \frac{1}{M_{Pl}^2}$ through the Newton constant $G$ and the
reduced Planck mass $M_{Pl}$, $R$ the scalar curvature, $\alpha_R, \alpha_C$ dimensionless positive coupling constants, 
and $C_{\mu\nu\rho\sigma}$ is conformal tensor defined as
\begin{eqnarray}
C_{\mu\nu\rho\sigma} &=& R_{\mu\nu\rho\sigma} - \frac{1}{2} ( g_{\mu\rho} R_{\nu\sigma}
- g_{\mu\sigma} R_{\nu\rho} - g_{\nu\rho} R_{\mu\sigma} + g_{\nu\sigma} R_{\mu\rho} )
\nonumber\\
&+& \frac{1}{6} ( g_{\mu\rho} g_{\nu\sigma} - g_{\mu\sigma} g_{\nu\rho} ) R.
\label{C-tensor}  
\end{eqnarray}

Since the second and third terms on the right-hand side (RHS) of (\ref{Class-Lag1}) include
higher-derivative terms, it is necessary to rewrite them into the first-order form for the canonical quantization by introducing 
an auxiliary, symmetric tensor field $K_{\mu\nu} = K_{\nu\mu}$ and a St\"{u}ckelberg-like vector field $A_\mu$, which is needed to 
avoid the second-class constraint \cite{Kimura1, Kimura2, Kimura3}.  Then, the classical Lagrangian (\ref{Class-Lag1}) can be 
rewritten as
\begin{eqnarray}
{\cal L}_c = \sqrt{-g} \left[ \frac{1}{2 \kappa^2} R + \gamma G_{\mu\nu} K^{\mu\nu} 
+ \beta_1 ( K_{\mu\nu} - \nabla_\mu A_\nu - \nabla_\nu A_\mu )^2 + \beta_2 ( K - 2 \nabla_\rho A^\rho )^2 \right],
\label{Class-Lag2}  
\end{eqnarray}
where $G_{\mu\nu} \equiv R_{\mu\nu} - \frac{1}{2} g_{\mu\nu} R$ denotes the Einstein tensor,
and $\gamma, \beta_1$ and $\beta_2$ are dimensionless coupling constants which obey a relation:
\begin{eqnarray}
\alpha_C = \frac{\gamma^2}{8 \beta_1},  \qquad
\alpha_R = - \frac{( \beta_1 + \beta_2 ) \gamma^2}{12 \beta_1 ( \beta_1 + 4 \beta_2 )}.
\label{Couplings}  
\end{eqnarray}
Here we assume that  $\beta_1 > 0, \beta_1 \neq - \beta_2, \beta_1 \neq 0, \beta_1 + 4 \beta_2 \neq 0$. 
In case of $\beta_1 = - \beta_2$, we have the well-known conformal gravity \cite{Oda-Ohta, Oda-Conf}.
   
The classical Lagrangian ${\cal L}_c$ is invariant under both general coordinate transformation (GCT)
and the St\"{u}ckelberg transformation (ST). The infinitesimal GCT is described as
\begin{eqnarray}
&{}& \delta^{(1)} g_{\mu\nu} = - ( \nabla_\mu \xi_\nu + \nabla_\nu \xi_\mu )
= - ( \xi^\alpha \partial_\alpha g_{\mu\nu} + \partial_\mu \xi^\alpha g_{\alpha\nu} 
+ \partial_\nu \xi^\alpha g_{\alpha\mu} ), 
\nonumber\\
&{}& \delta^{(1)} K_{\mu\nu} = - \xi^\alpha \nabla_\alpha K_{\mu\nu} - \nabla_\mu \xi^\alpha K_{\alpha\nu}
- \nabla_\nu \xi^\alpha K_{\mu\alpha},
\nonumber\\
&{}& \delta^{(1)} A_\mu = - \xi^\alpha \nabla_\alpha A_\mu - \nabla_\mu \xi^\alpha A_\alpha.
\label{GCT}  
\end{eqnarray}
while the infinitesimal ST takes the form:
\begin{eqnarray}
\delta^{(2)} g_{\mu\nu} =  0, \qquad
\delta^{(2)} K_{\mu\nu} = \nabla_\mu \varepsilon_\nu + \nabla_\nu \varepsilon_\mu, 
\qquad
\delta^{(2)} A_\mu = \varepsilon_\mu.
\label{Stuckel}  
\end{eqnarray}
In the above, $\xi_\mu$ and $\varepsilon_\mu$ are infinitesimal transformation parameters.

Next let us turn our attention to the construction of quantum quadratic gravity. 
For this purpose, we have to introduce the gauge fixing conditions for general coordinate transformation (GCT) and 
St\"{u}ckelberg transformation (ST). The appropriate gauge fixing condition 
for the GCT, which preserves the maximal global symmetry, i.e. the general linear transformation $GL(4)$ is provided by 
the de Donder gauge condition (or the harmonic gauge condition)\footnote{The $GL(4)$
transformation is needed to show that the graviton is an exactly massless Nambu-Goldstone boson associated with 
spontaneous symmetry breakdown from the $GL(4)$ to the Lorentz group $SO(1, 3)$ \cite{NO-grav} and it foribits a background dependent
term $\eta^{\mu\nu} B_\mu B_\nu$ to appear in the gauge fixing term.}:
\begin{eqnarray}
\partial_\mu \tilde g^{\mu\nu} = 0,
\label{Donder}  
\end{eqnarray}
where we have defined $\tilde g^{\mu\nu} \equiv \sqrt{-g} g^{\mu\nu}$.  For the ST, 
we impose the covariant gauge condition:
\begin{eqnarray}
\nabla_\mu K^{\mu\nu} = 0,
\label{K-gauge}  
\end{eqnarray}
which is also invariant under the $GL(4)$.  

After gauge fixing, the Lagrangian is not invariant under the local gauge transformations any longer, but it is invariant
under the corresponding BRST transformations, which are obtained from classical transformations by replacing the infinitesimal transformation 
parameters $\xi_\mu$ and $\varepsilon_\mu$ with Faddeev-Popov (FP) ghosts $c_\mu$ and $\zeta_\mu$, respectively. 
The BRST transformation for the GCT, which we call GCT BRST transformation, is then given by
\begin{eqnarray}
&{}& \delta^{(1)}_B g_{\mu\nu} = - ( \nabla_\mu c_\nu + \nabla_\nu c_\mu )
= - ( c^\alpha \partial_\alpha g_{\mu\nu} + \partial_\mu c^\alpha g_{\alpha\nu} 
+ \partial_\nu c^\alpha g_{\alpha\mu} ), 
\nonumber\\
&{}& \delta^{(1)}_B K_{\mu\nu} = - c^\alpha \nabla_\alpha K_{\mu\nu} - \nabla_\mu c^\alpha K_{\alpha\nu}
- \nabla_\nu c^\alpha K_{\mu\alpha}, 
\nonumber\\
&{}& \delta^{(1)}_B A_\mu = - c^\alpha \nabla_\alpha A_\mu - \nabla_\mu c^\alpha A_\alpha, \qquad
\delta^{(1)}_B c^\mu = - c^\alpha \partial_\alpha c^\mu,
\nonumber\\
&{}& \delta^{(1)}_B \bar c_\mu = i B_\mu, \qquad
\delta^{(1)}_B B_\mu = 0, \qquad
\delta^{(1)}_B b_\mu = - c^\alpha \partial_\alpha b_\mu,
\label{GCT-BRST}  
\end{eqnarray}
where $\bar c_\mu$ and $B_\mu$ are respectively an antighost and a Nakanishi-Lautrup (NL) field, and
a new NL field $b_\mu$ is defined as
\begin{eqnarray}
b_\mu = B_\mu - i c^\alpha \partial_\alpha \bar c_\mu,
\label{new-b}  
\end{eqnarray}
which will be used in place of $B_\mu$ in what follows.
The BRST transformation for the ST, which we call ST BRST transformation, 
is of form:
\begin{eqnarray}
&{}& \delta^{(2)}_B g_{\mu\nu} =  0, \qquad
\delta^{(2)}_B K_{\mu\nu} = \nabla_\mu \zeta_\nu + \nabla_\nu \zeta_\mu, 
\nonumber\\
&{}& \delta^{(2)}_B A_\mu = \zeta_\mu,  \qquad
\delta^{(2)}_B \bar \zeta_\mu = i \beta_\mu,  \qquad
\delta^{(2)}_B \zeta_\mu = \delta^{(2)}_B \beta_\mu = 0,
\label{STT-BRST}  
\end{eqnarray}
where $\bar \zeta_\mu$ and $\beta_\mu$ are respectively an antighost and a Nakanishi-Lautrup (NL) field. 
It is obvious that the two BRST transformations are nilpotent, $(\delta^{(1)}_B)^2 = (\delta^{(2)}_B)^2 = 0$.  
In order to make the two BRST transformations be anticommuting with each other, i.e.
$\{ \delta^{(1)}_B, \delta^{(2)}_B \} = 0$, the remaining BRST transformations are fixed to
\begin{eqnarray}
&{}& \delta^{(1)}_B \zeta_\mu = - c^\alpha \nabla_\alpha \zeta_\mu - \nabla_\mu c^\alpha \zeta_\alpha, \qquad
\delta^{(1)}_B \bar \zeta_\mu = - c^\alpha \nabla_\alpha \bar \zeta_\mu - \nabla_\mu c^\alpha \bar \zeta_\alpha,
\nonumber\\
&{}& \delta^{(1)}_B \beta_\mu = - c^\alpha \nabla_\alpha \beta_\mu - \nabla_\mu c^\alpha \beta_\alpha, \qquad
\delta^{(2)}_B b_\mu = \delta^{(2)}_B c^\mu = \delta^{(2)}_B \bar c_\mu = 0.
\label{Remain-BRST}  
\end{eqnarray}
 
Now that we have selected suitable gauge fixing conditions and established all the BRST transformations, 
it is straightforward to make a gauge fixed, BRST invariant quantum Lagrangian by following the standard recipe:
\begin{eqnarray}
{\cal L}_q &\equiv& {\cal L}_c + i \delta_B^{(1)} ( \tilde g^{\mu\nu} \partial_\mu \bar c_\nu ) 
+ i \delta_B^{(2)} ( \sqrt{-g} \, \bar \zeta_\nu \nabla_\mu K^{\mu\nu} )
\nonumber\\
&=& \sqrt{-g} \Bigl[ \frac{1}{2 \kappa^2} R + \gamma G_{\mu\nu} K^{\mu\nu} 
+ \beta_1 ( K_{\mu\nu} - \nabla_\mu A_\nu - \nabla_\nu A_\mu )^2 + \beta_2 ( K - 2 \nabla_\rho A^\rho )^2 \Bigr]
\nonumber\\
&-& \tilde g^{\mu\nu} \partial_\mu b_\nu - i \tilde g^{\mu\nu} \partial_\mu \bar c_\rho \partial_\nu c^\rho
+ \sqrt{-g} [ - \nabla_\mu K^{\mu\nu} \beta_\nu + i \nabla^\mu \bar  \zeta^\nu ( \nabla_\mu \zeta_\nu 
+ \nabla_\nu \zeta_\mu ) ],
\label{Quant-Lag}  
\end{eqnarray}
where surface terms are dropped.

Starting with the quantum Lagrangian (\ref{Quant-Lag}), it is straightforward to derive field equations by taking the variation
of each fundamental field in order whose result is given by
\begin{eqnarray}
&{}& \hat K_{\mu\nu} + \frac{\beta_2}{\beta_1} g_{\mu\nu} \hat K = - \frac{\gamma}{2 \beta_1} G_{\mu\nu} 
- \frac{1}{2 \beta_1} \nabla_{(\mu} \beta_{\nu)}. 
\label{Field-eq1}
\\
&{}& \nabla^\mu ( \hat K_{\mu\nu} + \frac{\beta_2}{\beta_1} g_{\mu\nu} \hat K ) = 0. 
\label{Field-eq2}
\\
&{}& \partial_\mu \tilde g^{\mu\nu} = 0.
\label{Field-eq3}
\\
&{}& g^{\mu\nu} \partial_\mu \partial_\nu c^\rho = g^{\mu\nu} \partial_\mu \partial_\nu \bar c_\rho = 0.
\label{Field-eq4}
\\
&{}& \nabla_\mu K^{\mu\nu} = 0.
\label{Field-eq5}
\\
&{}& \nabla^\mu ( \nabla_\mu \zeta_\nu + \nabla_\nu \zeta_\mu ) = \nabla^\mu ( \nabla_\mu \bar \zeta_\nu 
+ \nabla_\nu \bar \zeta_\mu ) = 0.
\label{Field-eq6}
\end{eqnarray}
In the above, we have omitted to write down field equation with respect to the metric since it is complicated and is not used later, 
and defined the following quantities:
\begin{eqnarray}
\hat K_{\mu\nu} &=& K_{\mu\nu} - \nabla_\mu A_\nu - \nabla_\nu A_\mu,  \qquad
\hat K = g^{\mu\nu} \hat K_{\mu\nu} = K - 2 \nabla_\rho A^\rho,
\nonumber\\
A_{(\mu} B_{\nu)} &\equiv& \frac{1}{2} ( A_\mu B_\nu + A_\nu B_\mu ).
\label{Various-defs}
\end{eqnarray}
The GCT BRST transformation of the field equation for $\bar c_\rho$ in (\ref{Field-eq5}) 
enables us to derive the field equation for $b_\rho$ \cite{Oda-Saake}:
\begin{eqnarray}
g^{\mu\nu} \partial_\mu \partial_\nu b_\rho = 0.
\label{b-rho-eq}  
\end{eqnarray}

We are now ready to perform the canonical quantization procedure.
To do that, let us first derive the concrete expressions of canonical conjugate momenta and set up the canonical 
(anti)commutation relations (CCRs), which will be used in evaluating various equal-time (anti)commutation relations (ETCRs). 
To simplify various expressions, we obey the following abbreviations \cite{N-O-text}:
\begin{eqnarray}
[ A, B^\prime ] &=& [ A(x), B(x^\prime) ] |_{x^0 = x^{\prime 0}},
\qquad \delta^3 = \delta(\vec{x} - \vec{x}^\prime), 
\nonumber\\
\tilde f &=& \frac{1}{\tilde g^{00}} = \frac{1}{\sqrt{-g} g^{00}},
\label{abbreviation}  
\end{eqnarray}
where we assume that $\tilde g^{00}$ is invertible. 

To remove second-order derivatives of the metric tensor involved in $R$ and $G_{\mu\nu}$, and regard $b_\mu$
and $\beta_\mu$ as non-canonical variables, by performing the integration by parts and neglecting surface terms 
the Lagrangian (\ref{Quant-Lag}) is cast to the form:
\begin{eqnarray}
{\cal L}_q &=& - \frac{1}{2 \kappa^2} \tilde g^{\mu\nu} ( \Gamma^\alpha_{\sigma\alpha} \Gamma^\sigma_{\mu\nu} 
- \Gamma^\alpha_{\sigma\nu} \Gamma^\sigma_{\mu\alpha} ) 
- \gamma \sqrt{-g} ( \Gamma^\alpha_{\mu\nu} \partial_\alpha - \Gamma^\alpha_{\mu\alpha} \partial_\nu
+ \Gamma^\beta_{\mu\alpha} \Gamma^\alpha_{\beta\nu} - \Gamma^\alpha_{\mu\alpha} \Gamma^\beta_{\nu\beta} )
\bar K^{\mu\nu} 
\nonumber\\
&+& \beta_1 \sqrt{-g} ( K_{\mu\nu} - \nabla_\mu A_\nu - \nabla_\nu A_\mu )^2 + \beta_2 \sqrt{-g} ( K - 2 \nabla_\mu A^\mu )^2  
+ \partial_\mu \tilde g^{\mu\nu} b_\nu
\nonumber\\
&-&  i \tilde g^{\mu\nu} \partial_\mu \bar c_\rho \partial_\nu c^\rho - \sqrt{-g} \, \nabla_\mu K^{\mu\nu} \cdot \beta_\nu 
+ i \sqrt{-g} \, \nabla^\mu \bar \zeta^\nu ( \nabla_\mu \zeta_\nu + \nabla_\nu \zeta_\mu ),
\label{Can-Quant-Lag}  
\end{eqnarray}
where $\bar K_{\mu\nu}$ is defined as
\begin{eqnarray}
\bar K_{\mu\nu} \equiv K_{\mu\nu} - \frac{1}{2} g_{\mu\nu} K,  \qquad
\bar K \equiv g^{\mu\nu} \bar K_{\mu\nu} = - K.
\label{bar-K}  
\end{eqnarray}
Since the NL fields $b_\mu$ and $\beta_\mu$ have no derivatives in ${\cal L}_q$, we can regard them as non-canonical 
variables.

Since the Lagrangian (\ref{Can-Quant-Lag}) is now written in the first-order formalism, it is straightforward to derive 
the concrete expressions of canonical conjugate momenta. The result is given by
\begin{eqnarray}
\pi_g^{\mu\nu} &=& \frac{\partial {\cal L}_q}{\partial \dot g_{\mu\nu}}, 
\nonumber\\
\pi_K^{\mu\nu} &=& \frac{\partial {\cal L}_q}{\partial \dot{K}_{\mu\nu}} 
= - \gamma \sqrt{-g} \, \Bigl[ ( g^{\mu\rho} g^{\nu\sigma} - \frac{1}{2} g^{\mu\nu} g^{\rho\sigma} ) 
\Gamma^0_{\rho\sigma}
- \frac{1}{2} ( g^{0\mu} g^{\nu\rho} + g^{0\nu} g^{\mu\rho} - g^{\mu\nu} g^{0\rho} ) \Gamma^\sigma_{\rho\sigma} \Bigr]
\nonumber\\ 
&-& \frac{1}{2} \sqrt{-g} \, ( g^{0\mu} \beta^\nu + g^{0\nu} \beta^\mu ),
\nonumber\\ 
\pi_A^\mu &=& \frac{\partial {\cal L}_q}{\partial \dot{A}_\mu} = - 4 \sqrt{-g} \, ( \beta_1 \hat K^{0\mu} + \beta_2 g^{0\mu} \hat K ),
\nonumber\\
\pi_{c \mu} &=& \frac{\partial {\cal L}_q}{\partial \dot c^\mu} = - i \tilde g^{0 \nu} \partial_\nu \bar c_\mu,  \qquad
\pi_{\bar c}^\mu = \frac{\partial {\cal L}_q}{\partial \dot {\bar c}_\mu} = i \tilde g^{0 \nu} \partial_\nu c^\mu,
\nonumber\\
\pi_\zeta^\mu &=& \frac{\partial {\cal L}_q}{\partial \dot \zeta_\mu} = i \sqrt{-g} \, ( \nabla^\mu \bar \zeta^0
+ \nabla^0 \bar \zeta^\mu ),  \qquad
\pi_{\bar \zeta}^\mu = \frac{\partial {\cal L}_q}{\partial \dot {\bar \zeta}_\mu} = - i \sqrt{-g} \, ( \nabla^\mu \zeta^0
+ \nabla^0 \zeta^\mu ),
\label{CCM}  
\end{eqnarray}
where we omit to write the expression of $\pi_g^{\mu\nu}$ explicitly, the dot stands for the derivative with respect to time, 
e.g. $\dot g_{\mu\nu} \equiv \frac{\partial g_{\mu\nu}}{\partial t} \equiv \frac{\partial g_{\mu\nu}}{\partial x^0} \equiv \partial_0 g_{\mu\nu}$, 
and the differentiation of ghosts is taken from the right.  
We are now able to set up the canonical (anti)commutation relations (CCRs): 
\begin{eqnarray}
&{}& [ g_{\mu\nu}, \pi_g^{\prime\rho\lambda} ] = [ K_{\mu\nu}, \pi_K^{\prime\rho\lambda} ] 
= i \frac{1}{2} ( \delta_\mu^\rho\delta_\nu^\lambda 
+ \delta_\mu^\lambda\delta_\nu^\rho) \delta^3,   \qquad
[ A_\mu, \pi_A^{\prime\nu} ] = i \delta_\mu^\nu \delta^3, 
\nonumber\\
&{}& \{ c^\mu, \pi_{c \nu}^\prime \} = \{ \bar c_\nu, \pi_{\bar c}^{\prime\mu} \}
= \{ \bar{\zeta}_\nu, \pi_{\bar \zeta}^{\prime\mu} \} = \{ \zeta_\nu, \pi_\zeta^{\prime\mu} \} = i \delta_\nu^\mu \delta^3,      
\label{CCRs}  
\end{eqnarray}
where the other (anti)commutation relations vanish. 

Finally, we can in principle calculate all the equal-time (anti)commutation relations (ETCRs) explicitly in closed form
by using the CCRs, field equations and BRST transformations.  One of the remarkable features in
quadratic gravity is that the ETCRs among the time derivatives of the metric tensor identically vanish:
\begin{eqnarray}
\left[ \frac{\partial^m g_{\rho\sigma}}{\partial t^m}, \frac{\partial^n g_{\mu\nu}^\prime}{\partial t^n} \right] = 0,
\label{zero-ETCR}  
\end{eqnarray}
where $m, n = 0, 1, 2, \dots$.
For instance, let us prove the first nontrivial case, $m= 0, n = 1$.
From the canonical conjugate momentum $\pi_K^{\mu\nu}$, $\dot g_{ij}$ can be described as
\begin{eqnarray}
\dot g_{ij} &=& \frac{2}{\gamma} \tilde f \Big[ ( g_{i\mu} g_{j\nu} - \frac{1}{2} g_{ij} g_{\mu\nu} ) 
\pi_K^{\mu\nu} - \frac{1}{2} \tilde g_{ij} \beta^0 \Big]
+ \tilde f [ \tilde g^{0\rho} ( \partial_i g_{j\rho} + \partial_j g_{i\rho} ) 
\nonumber\\
&-& \tilde g^{0k} \partial_k g_{ij} ].
\label{dot-g-piK}  
\end{eqnarray}
This expression immediately gives us the ETCR:
\begin{eqnarray}
[ \dot g_{ij}, g_{\mu\nu}^\prime ] = 0.
\label{g-dot-g}  
\end{eqnarray}
Here we have used the ETCR:
\begin{eqnarray}
[ \beta_\rho, g_{\mu\nu}^\prime ] = 0,
\label{beta-g}  
\end{eqnarray}
which can be easily shown by taking the ST BRST transformation (\ref{STT-BRST})
of the CCR, $[ \bar \zeta_\rho, g_{\mu\nu}^\prime ] = 0$. 

For the purpose of calculating the remaining ETCR, $[ \dot g_{0\rho}, g_{\mu\nu}^\prime ]$ we utilize 
the de Donder gauge condition (\ref{Donder}), from which $\dot g_{0\rho}$ can be expressed 
in terms of $\dot g_{ij}$:
\begin{eqnarray}
\dot g_{00} &=& \tilde f ( \tilde g^{ij} \dot g_{ij} - 2 \tilde g^{i\rho} \partial_i g_{0\rho} ),
\nonumber\\
\dot g_{0i} &=& \tilde f \Big( - \tilde g^{0j} \dot g_{ij} + \frac{1}{2} \tilde g^{\rho\sigma} \partial_i g_{\rho\sigma} 
- \tilde g^{j\rho} \partial_j g_{\rho i} \Big).
\label{dot-g-0rho}  
\end{eqnarray}
Then, using Eq. (\ref{g-dot-g}), it is easy to see that $[ \dot g_{0\rho}, g_{\mu\nu}^\prime ] = 0$, so we have a surprisingly 
simple ETCR: 
\begin{eqnarray}
[ \dot g_{\rho\sigma}, g_{\mu\nu}^\prime ] = 0.
\label{Zero ETCR}  
\end{eqnarray}
In a similar manner, we can prove the general formula (\ref{zero-ETCR}).

To close this review section, let us comment on physical states of quadratic gravity on the basis of the 
BRST formalism.  One can analyze asymptotic fields under the assumption that all elementary fields have their own
asymptotic fields and there is no bound state. We also assume that all asymptotic fields are
governed by the quadratic part of the quantum Lagrangian apart from possible renormalization.

Let us expand the gravitational field $g_{\mu\nu}$ around a flat Minkowski metric $\eta_{\mu\nu}$ 
as
\begin{eqnarray}
g_{\mu\nu} = \eta_{\mu\nu} + \varphi_{\mu\nu},
\label{Background}  
\end{eqnarray}
where $\varphi_{\mu\nu}$ denotes fluctuations. 
Then, up to surface terms the quadratic part of the quantum Lagrangian (\ref{Quant-Lag}) reads\footnote{For the sake of simplicity, 
we use the same notation for the other asymptotic fields as that for the interacting fields. }:
\begin{align}
&{\cal L}_q = \frac{1}{2 \kappa^2} \Bigg( \frac{1}{4} \varphi_{\mu\nu} \Box \varphi^{\mu\nu} 
- \frac{1}{4} \varphi \Box \varphi - \frac{1}{2} \varphi^{\mu\nu} \partial_\mu \partial_\rho \varphi_\nu{}^\rho
+ \frac{1}{2} \varphi^{\mu\nu} \partial_\mu \partial_\nu \varphi \Bigg )
\nonumber\\
&+ \gamma \left[ \left(\partial_\mu \partial_\rho \varphi_\nu{}^\rho - \frac{1}{2} \Box \varphi_{\mu\nu}
- \frac{1}{2} \partial_\mu \partial_\nu \varphi \right) K^{\mu\nu} + \frac{1}{2} ( \Box \varphi 
- \partial_\mu \partial_\nu \varphi^{\mu\nu} ) K \right]
\nonumber\\
&+ \beta_1 ( K_{\mu\nu} - \partial_\mu A_\nu - \partial_\nu A_\mu )^2
+ \beta_2 ( K - 2 \partial_\rho A^\rho )^2 + \left( \varphi^{\mu\nu} - \frac{1}{2} \eta^{\mu\nu} \varphi \right)
\partial_\mu b_\nu
\nonumber\\
& - i \partial_\mu \bar c_\rho \partial^\mu c^\rho - \partial_\mu K^{\mu\nu} \beta_\nu
+ i \partial^\mu \bar \zeta^\nu ( \partial_\mu \zeta_\nu + \partial_\nu \zeta_\mu). 
\label{Free-Lag}  
\end{align}
In this section, the spacetime indices $\mu, \nu, \dots$ are raised or lowered by the Minkowski metric $\eta^{\mu\nu}
= \eta_{\mu\nu} = \rm{diag} ( -1, 1, 1, 1)$, and we define $\Box \equiv \eta^{\mu\nu} \partial_\mu \partial_\nu$
and $\varphi \equiv \eta^{\mu\nu} \varphi_{\mu\nu}$.

From this Lagrangian, it is straightforward to derive the following linearized field equations: 
\begin{eqnarray}
&{}& \Box b_\mu = \Box c^\mu = \Box \bar c_\mu = 0, \qquad
\Box^2 \beta_\mu = \Box^2 \zeta_\mu = \Box^2 \bar \zeta_\mu = 0, \qquad
\nonumber\\
&{}&  (\Box - m^2 ) \phi = \Box^2 \tilde A_\mu = 0, \qquad
( \Box - M^2 ) \psi_{\mu\nu} = \partial^\mu \psi_{\mu\nu} = \eta^{\mu\nu} \psi_{\mu\nu} = 0,
\nonumber\\
&{}& \Box^2 h_{\mu\nu} = \partial^\mu h_{\mu\nu} - \frac{1}{2} \partial_\nu h = 0,
\label{Lin-field-eqs}  
\end{eqnarray}
where we have defined
\begin{eqnarray}
\phi &=& \hat K - \frac{\kappa^2 \gamma}{\beta_1 + 4 \beta_2} \partial_\rho b^\rho
+ \frac{1}{2 (\beta_1 + 4 \beta_2)} \partial_\rho \beta^\rho,  \qquad  
\tilde A_\mu = A_\mu - \frac{\beta_2}{2 \beta_1 m^2} \partial_\mu \phi,
\nonumber\\
\psi_{\mu\nu} &=& \hat K_{\mu\nu} - \frac{\beta_1 + \beta_2}{3 \beta_1} \left( \eta_{\mu\nu} 
+ \frac{\kappa^2 \gamma^2}{\beta_1} \partial_\mu \partial_\nu \right) \phi
+ \frac{\kappa^2 \gamma}{\beta_1} \Bigg[ \partial_{(\mu} b_{\nu)} 
- \frac{\beta_1 + 2 \beta_2}{2 (\beta_1 + 4 \beta_2)} \eta_{\mu\nu} \partial_\rho b^\rho
\nonumber\\
&-& \frac{(\beta_1 + 2 \beta_2) \kappa^2 \gamma^2}{2 \beta_1 (\beta_1 + 4 \beta_2)} 
\partial_\mu \partial_\nu \partial_\rho b^\rho \Bigg]
+ \frac{1}{2 \beta_1} \Bigg[ \partial_{(\mu} \beta_{\nu)} 
- \frac{\beta_2}{\beta_1 + 4 \beta_2} \eta_{\mu\nu} \partial_\rho \beta^\rho
\nonumber\\
&-& \frac{\beta_2 \kappa^2 \gamma^2}{\beta_1 (\beta_1 + 4 \beta_2)} 
\partial_\mu \partial_\nu \partial_\rho \beta^\rho \Bigg],
\nonumber\\
h_{\mu\nu} &=& \varphi_{\mu\nu} - 2 \kappa^2 \gamma \psi_{\mu\nu} 
- \frac{2 ( \beta_1 + \beta_2 ) \kappa^2 \gamma}{3 \beta_1} \Bigg( \eta_{\mu\nu} 
+ \frac{2}{m^2} \partial_\mu \partial_\nu \Bigg) \phi,
\nonumber\\
m^2 &=& - \frac{\beta_1 ( \beta_1 + 4 \beta_2 )}{( \beta_1 + \beta_2 ) \kappa^2 \gamma^2},  \qquad
M^2 = \frac{2 \beta_1}{\kappa^2 \gamma^2}.
\label{mass-square}  
\end{eqnarray}
From the BRST quartet mechanism \cite{Kugo-Ojima}, it turns out that physical states of quadratic gravity are composed of
the spin $0$ massive scalar $\phi$, the spin $2$ massless graviton $h_{\mu\nu}$ and the spin $2$ massive 
ghost $\psi_{\mu\nu}$. In particular, the massive ghost has negative norm so it breaks the unitarity of the physical 
S-matrix \cite{Oda-Can}.

\section{Higher-derivative de Donder gauge}

In case of gravitational theories with higher-derivative terms, it is known that we need to introduce ``third''
ghost in addition to the conventional FP ghosts at one-loop level in the path integral formalism \cite{Barth} and
it was later understood at full order on the basis of the operator formalism, or equivalently the BRST formalism \cite{Ohta}.
It is thus natural to introduce the higher-derivative generalization of the de Donder gauge condition for general coordinate 
transformation (GCT) in quadratic gravity and ask ourselves if the result of vanishing equal-time commutation relations 
of the metric tensor is still valid in such a higher-derivative gauge condition or not.  We will see that this is indeed the case. 
This fact implies that the vanishing ETCRs of the metric field hold true when we confine ourselves to the Donder gauge and its
higher-dimensional generalization.

Let us start with the construction of higher-derivative de Donder gauge based on the BRST formalism. Following the standard recipe,
the gauge fixing and FP ghost term can be made by a BRST-exact form:
\begin{eqnarray}
{\cal{L}}_{GF+FP}^{(1)} = i \delta_B^{(1)} ( \bar c_\mu A^\mu{}_\nu F^\nu ),
\label{High-Donder}  
\end{eqnarray}
where we have defined\footnote{It might be of interest to consider the generalized de Donder gauge, $\partial_\mu \tilde g^{\mu\nu} 
+ \alpha_0 \eta^{\mu\nu} B_\mu = 0$ with $\alpha_0$ being a gauge parameter although the term $\alpha_0 \eta^{\mu\nu} B_\mu$
is prohibited by the $GL(4)$ invariance.} 
\begin{eqnarray}
A_{\mu\nu} &=& a_1 g_{\mu\nu} \Box + a_2 \nabla_\mu \nabla_\nu + a_3 \nabla_\nu \nabla_\mu + a_4 R_{\mu\nu} 
+ a_5 g_{\mu\nu} R,
\nonumber\\
F^\nu &=& \partial_\mu \tilde g^{\mu\nu},
\label{A-F}  
\end{eqnarray}
where $a_i (i = 1, \dots, 5)$ are constants \cite{Ohta, Buchbinder}. Owing to the metric condition $\nabla_\mu g_{\nu\rho} = 0$, 
$A^\mu{}_\nu F^\nu$ can be cast to the form:
\begin{eqnarray}
A^\mu{}_\nu F^\nu = a R^\mu{}_\nu \partial_\rho \tilde g^{\rho\nu} + b R \partial_\rho \tilde g^{\rho\mu},
\label{A-F-result}  
\end{eqnarray}
where we have redefined $a = a_4, b = a_5$. We will call $A^\mu{}_\nu F^\nu = 0$ ``the higher-derivative de Donder
gauge'' in what follows. In case of gauge fixing condition for the St\"{u}ckelberg transformation (ST), one can slightly
extend the covariant gauge (\ref{K-gauge}) by adding a term proportional to the St\"{u}ckelberg field like
\begin{eqnarray}
\nabla_\mu K^{\mu\nu} + c A^\nu = 0,
\label{Ext-K-gauge}  
\end{eqnarray}
where $c$ is a constant. With this gauge condition, the gauge fixing and FP ghost term is given by
\begin{eqnarray}
{\cal{L}}_{GF+FP}^{(2)} = i \delta_B^{(2)} [ \sqrt{-g} \, \bar \zeta_\nu ( \nabla_\mu K^{\mu\nu} + c A^\nu ) ].
\label{Ext-K-gauge2}  
\end{eqnarray}
Thus, we can obtain the gauge fixed, BRST invariant quantum Lagrangian:  
\begin{eqnarray}
{\cal L}_q &\equiv& {\cal L}_c + i \delta_B^{(1)} ( \bar c_\mu A^\mu{}_\nu F^\nu ) 
+  i \delta_B^{(2)} [ \sqrt{-g} \, \bar \zeta_\nu ( \nabla_\mu K^{\mu\nu} + c A^\nu ) ]
\nonumber\\
&=& \sqrt{-g} \Bigl[ \frac{1}{2 \kappa^2} R + \gamma G_{\mu\nu} K^{\mu\nu} 
+ \beta_1 ( K_{\mu\nu} - \nabla_\mu A_\nu - \nabla_\nu A_\mu )^2 + \beta_2 ( K - 2 \nabla_\rho A^\rho )^2 \Bigr]
\nonumber\\
&-& B_\mu ( a R^\mu{}_\nu \partial_\rho \tilde g^{\rho\nu}  + b R \partial_\nu \tilde g^{\mu\nu} )
- i \bar c_\mu \Bigg\{ a \Bigg[ ( \nabla^\mu c^\rho + \nabla^\rho c^\mu ) R_{\rho\nu} + \nabla^\mu \nabla_\nu \nabla_\rho c^\rho
\nonumber\\
&+& \frac{1}{2} \Box ( \nabla^\mu c_\nu + \nabla_\nu c^\mu ) - \frac{1}{2} \nabla_\rho \nabla^\mu 
( \nabla^\rho c_\nu + \nabla_\nu c^\rho ) - \frac{1}{2} \nabla_\rho \nabla_\nu ( \nabla^\rho c^\mu + \nabla^\mu c^\rho ) \Bigg]
\nonumber\\
&-& b \delta_\nu^\mu \nabla_\rho R \, c^\rho \Bigg\}  \partial_\sigma \tilde g^{\sigma\nu}
- i \bar c_\mu ( a R^\mu{}_\nu + b \delta_\nu^\mu R ) \partial_\rho [ \sqrt{-g} ( \nabla^\rho c^\nu + \nabla^\nu c^\rho
- g^{\rho\nu} \nabla_\sigma c^\sigma ) ]
\nonumber\\
&+& \sqrt{-g} [ - ( \nabla_\mu K^{\mu\nu} + c A^\nu ) \beta_\nu + i \nabla^\mu \bar  \zeta^\nu ( \nabla_\mu \zeta_\nu 
+ \nabla_\nu \zeta_\mu ) - i c \bar \zeta_\mu \zeta^\mu ].
\label{Quant-Ext-Lag}  
\end{eqnarray}

We can derive all the field equations, but for later convenience we shall present only the field equation for $K^{\mu\nu}$:
\begin{eqnarray}
&{}& \hat K_{\mu\nu} + \frac{\beta_2}{\beta_1} g_{\mu\nu} \hat K = - \frac{\gamma}{2 \beta_1} G_{\mu\nu} 
- \frac{1}{2 \beta_1} \nabla_{(\mu} \beta_{\nu)},
\label{Ext-Field-eq}
\end{eqnarray}
which is the same as Eq. (\ref{Field-eq1}) in the de Donder gauge. Then, $R_{\mu\nu}$ and $R$ can 
be read off from Eq. (\ref{Ext-Field-eq}):
\begin{eqnarray}
R_{\mu\nu} &=& - \frac{2 \beta_1}{\gamma} \hat K_{\mu\nu} + \frac{\beta_1 + 2 \beta_2}{\gamma} g_{\mu\nu} \hat K 
- \frac{1}{\gamma} \left( \nabla_{(\mu} \beta_{\nu)} - \frac{1}{2} g_{\mu\nu} \nabla_\rho \beta^\rho \right),
\nonumber\\
R &=& \frac{2 (\beta_1 + 4 \beta_2 )}{\gamma} \hat K + \frac{1}{\gamma} \nabla_\rho \beta^\rho.
\label{Ext-Field-R}
\end{eqnarray}

\section{Vanishing equal-time commutation relation of metric}
    
In this section, we would like to show that in quadratic gravity all the ETCRs among the time derivatives of the metric 
tensor identically vanish in the higher-derivative de Donder gauge as in the conventional de Donder gauge.
The reason why we have the same result is that we make use of only the field equation and the canonical conjugate
momentum for the auxiliary field $K_{\mu\nu}$ which have the same form between the de Donder gauge and 
the higher-derivative de Donder gauge. A subtle point lies only in the difference between the de Donder gauge and 
the higher-derivative de Donder gauge. We shall carefully investigate a consequence of the higher-derivative de Donder gauge 
and verify that all the ETCRs among the time derivatives of the metric tensor identically vanish even in the higher-derivative 
de Donder gauge.

Along the same line of argument, we can show Eq. (\ref{g-dot-g}) since the expression of $\pi_K^{\mu\nu}$ is
the same in both the de Donder gauge and the higher-derivative de Donder gauge. Given Eq. (\ref{g-dot-g}), 
from the dimensional analysis and symmetry of indices, only the nonvanishing ETCR is of form
\begin{eqnarray}
[ \dot g_{\rho\sigma}, g_{\mu\nu}^\prime ] = \kappa^2 x_1 \delta_\rho^0 \delta_\sigma^0 \delta_\mu^0 \delta_\nu^0 
\delta^3 \equiv x \delta_\rho^0 \delta_\sigma^0 \delta_\mu^0 \delta_\nu^0 \delta^3, 
\label{Ext-dot-g-g}
\end{eqnarray}
where $x_1$ is a dimensionless constant. Here note that when we denote mass dimension of some quantity $\Phi$
as $[ \Phi ]$, we can see that $[ \dot g_{\rho\sigma} ] = M^1, [ g_{\mu\nu}^\prime ] = M^0, [ \kappa^2 ] = M^{-2}, [ \delta^3 ] = M^3$, 
so the both sides of this equation have the common dimension $M^1$ as required.

To fix the coefficient $x$, we appeal to the higher-derivative de Donder gauge, i.e. we require 
\begin{eqnarray}
[ ( a R^\lambda{}_\rho + b \delta_\rho^\lambda R ) \partial_\sigma \tilde g^{\rho\sigma}, g_{\mu\nu}^\prime ] = 0. 
\label{Hi-Donder-g}
\end{eqnarray}
If we define 
\begin{eqnarray}
I^\lambda{}_\rho \equiv a R^\lambda{}_\rho + b \delta_\rho^\lambda R, 
\label{I-lambda-rho}
\end{eqnarray}
this quantity in general includes the second derivative of the metric, i.e. $\partial^2 g$, but we can rewrite it in such a way 
that it includes only the first derivative of the metric, i.e. $\partial g$ by using field equation. Actually, Eq. (\ref{Ext-Field-R})
makes it possible to rewrite $I^\lambda{}_\rho$ into the form:
\begin{eqnarray}
I^\lambda{}_\rho &=& - \frac{2 a \beta_1}{\gamma} \hat K^\lambda{}_\rho + \frac{a (\beta_1 + 2 \beta_2 ) + 2b (\beta_1 + 4 \beta_2 )}{\gamma} 
\delta^\lambda{}_\rho \hat K 
- \frac{a}{\gamma} \nabla^{(\lambda} \beta_{\rho)} 
\nonumber\\
&+& \frac{a + 2b}{2 \gamma} \delta^\lambda{}_\rho \nabla_\sigma \beta^\sigma,
\label{I-lambda-rho2}
\end{eqnarray}
which involves at most the first derivative of the metric.

In order to evaluate the LHS of Eq. (\ref{Hi-Donder-g}), it is necessary to calculate $[ \Gamma^\lambda_{\rho\sigma}, g_{\mu\nu}^\prime ]$
whose result is given by
\begin{eqnarray}
[ \Gamma^\lambda_{\rho\sigma}, g_{\mu\nu}^\prime ] = \frac{1}{2} x g^{0\lambda} \delta_\rho^0 \delta_\sigma^0 \delta_\mu^0 \delta_\nu^0 \delta^3,
\label{Gamma-g}
\end{eqnarray}
where Eq. (\ref{Ext-dot-g-g}) was used. Moreover, one finds that
\begin{eqnarray}
[ \dot A_\rho, g_{\mu\nu}^\prime ] = \frac{1}{2} x \delta_\rho^0 \delta_\mu^0 \delta_\nu^0 \delta^3.
\label{dot-A-g}
\end{eqnarray}
A proof of this equation can be carried out as follows: Solving $\pi_A^\mu$ in (\ref{CCM}), which is common in both de Donder
gauges, in terms of $\dot A_\mu$ produces
\begin{eqnarray}
\dot A_0 &=& \frac{1}{2 (\beta_1 + \beta_2) g^{00}} \Bigg\{ - \frac{\beta_1 + 2 \beta_2}{\beta_1 g^{00}}  g^{0i} 
\Bigg [ \frac{1}{4 \sqrt{-g}} g_{i\mu} \pi_A^\mu + 2 \beta_1 g^{0\rho} \Gamma^\lambda_{\rho i} A_\lambda
\nonumber\\
&-& \beta_1 (  g^{0j} \partial_j A_i + g^{0\rho} \partial_i A_\rho ) + \beta_1 g^{0\rho} K_{\rho i} \Bigg]
+ \frac{1}{4 \sqrt{-g}} g_{0\mu} \pi_A^\mu + 2 ( \beta_1 g^{0\rho} \Gamma^\lambda_{0\rho} 
+ \beta_2 g^{\rho\sigma} \Gamma^\lambda_{\rho\sigma} ) A_\lambda 
\nonumber\\
&-& \beta_1 g^{0i} \partial_i A_0 - 2 \beta_2 g^{i\rho} \partial_i A_\rho + \beta_1 g^{0\rho} K_{0\rho} + \beta_2 K \Bigg\},
\nonumber\\
\dot A_i &=& \frac{1}{\beta_1 g^{00}} \Bigg [ \frac{1}{4 \sqrt{-g}} g_{i\mu} \pi_A^\mu + 2 \beta_1 g^{0\rho} \Gamma^\lambda_{\rho i} A_\lambda
- \beta_1 (  g^{0j} \partial_j A_i + g^{0\rho} \partial_i A_\rho ) + \beta_1 g^{0\rho} K_{\rho i} \Bigg].
\label{dot-A-exp}
\end{eqnarray}
With the help of Eq. (\ref{Gamma-g}), it turns out that these expressions lead to Eq. (\ref{dot-A-g}). 
Then, Eqs. (\ref{Gamma-g}) and (\ref{dot-A-g}) provide us with
\begin{eqnarray}
[ \nabla_\rho A_\sigma, g_{\mu\nu}^\prime ] = [ \hat K_{\rho\sigma}, g_{\mu\nu}^\prime ] = 0.
\label{nabla-A-K-g}
\end{eqnarray}

Next we will consider the CCR, $[ \pi_\zeta^\rho, g_{\mu\nu}^\prime ] = 0$. Using Eqs. (\ref{CCM}) and (\ref{Gamma-g}),
this CCR yields\footnote{Note that $\pi_\zeta^\rho$ takes the same expression in both the de Donder gauge and its higher-derivative one.} 
\begin{eqnarray}
[ \dot{\bar \zeta}_\rho, g_{\mu\nu}^\prime ] = \frac{1}{2} x \delta_\rho^0 \delta_\sigma^0 \delta_\mu^0 \delta_\nu^0 
\bar \zeta^0 \delta^3.
\label{dot-bar-Z-g}
\end{eqnarray}
Taking its ST BRST transformation leads to
\begin{eqnarray}
[ \dot \beta_\rho, g_{\mu\nu}^\prime ] = \frac{1}{2} x \delta_\rho^0 \delta_\sigma^0 \delta_\mu^0 \delta_\nu^0 \beta^0 \delta^3.
\label{dot-beta-g}
\end{eqnarray}
Together with Eq. (\ref{Gamma-g}), this equation gives us
\begin{eqnarray}
[ \nabla_\rho \beta_\sigma, g_{\mu\nu}^\prime ] = 0.
\label{nabla-beta-g}
\end{eqnarray}
Accordingly, we arrive at the result:
\begin{eqnarray}
[ I^\lambda{}_\rho, g_{\mu\nu}^\prime ] \equiv [ a R^\lambda{}_\rho + b \delta_\rho^\lambda R, g_{\mu\nu}^\prime ] =0. 
\label{I-g-prime}
\end{eqnarray}
Finally, since we can show that
\begin{eqnarray}
[ \partial_\sigma \tilde g^{\rho\sigma}, g_{\mu\nu}^\prime ] = - \sqrt{-g} g^{\alpha\beta} [ \Gamma^\rho_{\alpha\beta}, g_{\mu\nu}^\prime ] 
= - \frac{1}{2} x \tilde g^{00} g^{0\rho} \delta_\mu^0 \delta_\nu^0 \delta^3,
\label{de-Donder-g}
\end{eqnarray}
Eq. (\ref{Hi-Donder-g}) reads
\begin{eqnarray}
I^\lambda{}_\rho \times \left(- \frac{1}{2} \right) x \tilde g^{00} g^{0\rho} \delta_\mu^0 \delta_\nu^0 \delta^3 = 0. 
\label{Hi-Donder-dot-g2}
\end{eqnarray}
Hence, we can obtain $x =0$, which means the desired equation:\footnote{We present an alternative derivation 
without using field equation in Appendix A.}
\begin{eqnarray}
[ \dot g_{\rho\sigma}, g_{\mu\nu}^\prime ] = 0. 
\label{Ext-dot-g2}
\end{eqnarray}

We now proceed to a proof of the case:
\begin{eqnarray}
[ \dot g_{\rho\sigma}, \dot g_{\mu\nu}^\prime ] = 0. 
\label{Ext-dot-g-dot-g}
\end{eqnarray}
To do that, let us first notice that (\ref{dot-g-piK}) gives us
\begin{eqnarray}
[ \dot g_{ij}, \dot g_{kl}^\prime ] = 0,
\label{Ext-dot-g-dot-g2}
\end{eqnarray}
since we can easily show that $[ \beta_\mu, \pi_K^{\prime \rho\sigma} ] = [ \beta_\mu, \beta_\nu^\prime ] = 0$.
Furthermore, the dimensional analysis and an antisymmetric property, $[ \dot g_{\rho\sigma}, \dot g_{\mu\nu}^\prime ] = 
- [ \dot g_{\mu\nu}^\prime, \dot g_{\rho\sigma} ]$ enable us to write out a general expression:
\begin{eqnarray}
[ \dot g_{\rho\sigma}, \dot g_{\mu\nu}^\prime ] = \kappa^2 y_1 \delta_\rho^0 \delta_\sigma^0 \delta_\mu^0 \delta_\nu^0  
( g^{0k} \partial_k - g^{\prime 0k} \partial_k^\prime ) \delta^3
\equiv y (x, x^\prime) \delta_\rho^0 \delta_\sigma^0 \delta_\mu^0 \delta_\nu^0 \delta^3,
\label{Ext-dot-g-dot-g3}
\end{eqnarray}
where $y_1$ is a dimensionless constant and $y (x, x^\prime) \equiv \kappa^2 y_1 ( g^{0k} \partial_k - g^{\prime 0k} \partial_k^\prime )$.

To fix $y (x, x^\prime)$, let us impose a consistency condition with the higher-derivative de Donder gauge on
$\dot g_{\mu\nu}$:
\begin{eqnarray}
[ ( a R^\lambda{}_\rho + b \delta_\rho^\lambda R ) \partial_\sigma \tilde g^{\rho\sigma}, \dot g_{\mu\nu}^\prime ] 
\equiv [ I^\lambda{}_\rho \partial_\sigma \tilde g^{\rho\sigma}, \dot g_{\mu\nu}^\prime ]
= 0. 
\label{Hi-Donder-dot-g}
\end{eqnarray}
In evaluating this commutation relation, it is convenient to utilize the formula:
\begin{eqnarray}
R_{\mu\nu\rho\sigma} &=& \frac{1}{2} ( \partial_\mu \partial_\sigma g_{\nu\rho} + \partial_\nu \partial_\rho g_{\mu\sigma}
- \partial_\nu \partial_\sigma g_{\mu\rho} - \partial_\mu \partial_\rho g_{\nu\sigma} ) 
\nonumber\\
&+& g_{\alpha\beta} ( \Gamma^\alpha_{\mu\sigma} \Gamma^\beta_{\nu\rho} - \Gamma^\alpha_{\mu\rho} 
\Gamma^\beta_{\nu\sigma} ). 
\label{Riemann}
\end{eqnarray}
Then, by means of Eqs. (\ref{Ext-dot-g2}), (\ref{Ext-dot-g-dot-g3}) and (\ref{Riemann}), we find that 
\begin{eqnarray}
[ R^\lambda{}_\rho, \dot g_{\mu\nu}^\prime ] &=& \frac{1}{2} \Bigg\{ ( g^{\lambda\alpha} g^{0\beta} - g^{0\lambda} g^{\alpha\beta} )
\delta_\rho^0 \Big[  [ \ddot g_{\alpha\beta}, \dot g_{\mu\nu}^\prime ] + ( \delta_\alpha^0 \delta_\beta^i + \delta_\alpha^i \delta_\beta^0 )
\delta_\mu^0 \delta_\nu^0 \partial_i ( y \delta^3 )
\nonumber\\
&+& y \Gamma_{\alpha\beta}^0 \delta_\mu^0 \delta_\nu^0 \delta^3 \Big] + ( g^{0\lambda} g^{0\alpha} - g^{00} g^{\lambda\alpha} ) 
\Big[  [ \ddot g_{\alpha\rho}, \dot g_{\mu\nu}^\prime ] 
+ ( \delta_\alpha^0 \delta_\rho^i + \delta_\alpha^i \delta_\rho^0 ) \delta_\mu^0 \delta_\nu^0 \partial_i ( y \delta^3 )
\nonumber\\
&+& y \Gamma_{\alpha\rho}^0 \delta_\mu^0 \delta_\nu^0 \delta^3 \Big] \Bigg\}.
\label{Ricci-dot-g}
\end{eqnarray}
Using this equation, it is straightforward to calculate (\ref{Hi-Donder-dot-g}), which is given by
\begin{eqnarray}
&{}& \frac{1}{2} a \Bigg\{ ( g^{\lambda\alpha} g^{0\beta} - g^{0\lambda} g^{\alpha\beta} )
\delta_\rho^0 \Big[  [ \ddot g_{\alpha\beta}, \dot g_{\mu\nu}^\prime ] + ( \delta_\alpha^0 \delta_\beta^i + \delta_\alpha^i \delta_\beta^0 )
\delta_\mu^0 \delta_\nu^0 \partial_i ( y \delta^3 )
+ y \Gamma_{\alpha\beta}^0 \delta_\mu^0 \delta_\nu^0 \delta^3 \Big]
\nonumber\\
&+& ( g^{0\lambda} g^{0\alpha} - g^{00} g^{\lambda\alpha} ) \Big[  [ \ddot g_{\alpha\rho}, \dot g_{\mu\nu}^\prime ] 
+ ( \delta_\alpha^0 \delta_\rho^i + \delta_\alpha^i \delta_\rho^0 ) \delta_\mu^0 \delta_\nu^0 \partial_i ( y \delta^3 )
+ y \Gamma_{\alpha\rho}^0 \delta_\mu^0 \delta_\nu^0 \delta^3 \Big] \Bigg\} \partial_\sigma \tilde g^{\rho\sigma}
\nonumber\\
&+& b \delta_\rho^\lambda ( g^{0\alpha} g^{0\beta} - g^{00} g^{\alpha\beta} )
\Big[  [ \ddot g_{\alpha\beta}, \dot g_{\mu\nu}^\prime ] + ( \delta_\alpha^0 \delta_\beta^i + \delta_\alpha^i \delta_\beta^0 )
\delta_\mu^0 \delta_\nu^0 \partial_i ( y \delta^3 )
+ y \Gamma_{\alpha\beta}^0 \delta_\mu^0 \delta_\nu^0 \delta^3 \Big] \partial_\sigma \tilde g^{\rho\sigma}
\nonumber\\
&-& \frac{1}{2} ( a R^\lambda{}_\rho + b \delta_\rho^\lambda R ) \tilde g^{00} g^{0\rho} y \delta_\mu^0 \delta_\nu^0 \delta^3
= 0.
\label{Const-eq}
\end{eqnarray}

Next, let us write out an explicit expression of the ETCR, $[ \ddot g_{\rho\sigma}, \dot g_{\mu\nu}^\prime ]$ from a symmetry property.
From $[ \ddot g_{\rho\sigma}, \dot g_{\mu\nu}^\prime ] = [ \ddot g_{\mu\nu}^\prime, \dot g_{\rho\sigma} ]$, this ETCR is invariant 
under the simultaneous exchange of $(\mu, \nu) \leftrightarrow (\rho, \sigma)$ and primed $\leftrightarrow$ unprimed as well as 
the symmtery $\mu \leftrightarrow \nu$ and $\rho \leftrightarrow \sigma$. Also taking account of dimensional consideration leads to
\begin{eqnarray}
[ \ddot g_{\rho\sigma}, \dot g_{\mu\nu}^\prime ] &=& \kappa^2 [ z_1 ( \ddot g_{\rho\sigma} g_{\mu\nu} + g_{\rho\sigma} \ddot g_{\mu\nu} )
+ z_2 \dot g_{\rho\sigma} \dot g_{\mu\nu} + z_3 ( \dot g_{\rho\mu} \dot g_{\sigma\nu} + \dot g_{\rho\nu} \dot g_{\sigma\mu} ) 
\nonumber\\ 
&+& z_4 ( \delta_\rho^0 \delta_\sigma^0 \ddot g_{\mu\nu} + \delta_\mu^0 \delta_\nu^0 \ddot g_{\rho\sigma} )
+ z_5 ( \delta_\rho^0 \delta_\mu^0 \ddot g_{\sigma\nu} + \delta_\rho^0 \delta_\nu^0 \ddot g_{\sigma\mu} 
+ \delta_\sigma^0 \delta_\mu^0 \ddot g_{\rho\nu} + \delta_\sigma^0 \delta_\nu^0 \ddot g_{\rho\mu} )
\nonumber\\
&+& z_6 \delta_\rho^0 \delta_\sigma^0 \delta_\mu^0 \delta_\nu^0 \ddot g_{00} + z_7 \delta_\rho^0 \delta_\sigma^0 \delta_\mu^0 \delta_\nu^0 
(\dot g_{00})^2] \delta^3
\nonumber\\
&+& \kappa^2 z_8 ( \delta_\rho^0 \delta_\sigma^0 \dot g_{\mu\nu} g^{0k} - \delta_\mu^0 \delta_\nu^0 \dot g_{\rho\sigma}^\prime g^{\prime 0k} ) 
\partial_k \delta^3
+ \kappa^2 \Big\{ [ z_9 ( \delta_\rho^0 \delta_\mu^0 \dot g_{\rho\nu} + \delta_\rho^0 \delta_\nu^0 \dot g_{\sigma\mu} 
\nonumber\\
&+& \delta_\sigma^0 \delta_\mu^0 \dot g_{\rho\nu} + \delta_\sigma^0 \delta_\nu^0 \dot g_{\rho\mu} ) 
+ z_{10} \delta_\rho^0 \delta_\sigma^0 \delta_\mu^0 \delta_\nu^0 \dot g_{00}  ] g^{0k}
\nonumber\\
&-& ( {\rm{unprimed}}  \rightarrow {\rm{primed}} ) \Big\} \partial_k \delta^3.
\label{Form-ddot-g-dot-g}
\end{eqnarray}
Substituting this equation (\ref{Form-ddot-g-dot-g}) into (\ref{Const-eq}), we find that the only
nontrivial solution takes the form:
\begin{eqnarray}
&{}& y ( x, x^\prime ) = 0,
\nonumber\\
&{}& [ \ddot g_{\rho\sigma}, \dot g_{\mu\nu}^\prime ] = \kappa^2 [ z_6 \ddot g_{00} + z_7 (\dot g_{00} )^2 
+ z_{10} ( \dot g_{00} g^{0k} - \dot g_{00}^\prime g^{\prime 0k} ) \partial_k ] \delta_\rho^0 \delta_\sigma^0 
\delta_\mu^0 \delta_\nu^0 \delta^3 
\nonumber\\
&{}& \equiv z(x, x^\prime) \delta_\rho^0 \delta_\sigma^0 \delta_\mu^0 \delta_\nu^0 \delta^3,
\label{Nontrivial-sol}
\end{eqnarray}
where we have defined $z(x, x^\prime) \equiv \kappa^2 [ z_6 \ddot g_{00} + z_7 (\dot g_{00} )^2 
+ z_{10} ( \dot g_{00} g^{0k} - \dot g_{00}^\prime g^{\prime 0k} ) \partial_k ]$. The former equation 
clearly exhibits Eq. (\ref{Ext-dot-g-dot-g}).

In a perfectly similar manner, let us impose a consistency condition with the higher-derivative de Donder gauge on
$\ddot g_{\mu\nu}$
\begin{eqnarray}
[ ( a R^\lambda{}_\rho + b \delta_\rho^\lambda R ) \partial_\sigma \tilde g^{\rho\sigma}, \ddot g_{\mu\nu}^\prime ] 
\equiv [ I^\lambda{}_\rho \partial_\sigma \tilde g^{\rho\sigma}, \ddot g_{\mu\nu}^\prime ]
= 0,
\label{Hi-Donder-ddot-g}
\end{eqnarray}
and then consider a general expression of $[ \ddot g_{\rho\sigma}, \ddot g_{\mu\nu}^\prime ]$.  
Because of $[ \ddot g_{\rho\sigma}, \ddot g_{\mu\nu}^\prime ] = - [ \ddot g_{\mu\nu}^\prime, \ddot g_{\rho\sigma} ]$, it is antisymmetric
under the simultaneous exchange of $(\mu, \nu) \leftrightarrow (\rho, \sigma)$ and primed $\leftrightarrow$ unprimed as well as 
the symmtery $\mu \leftrightarrow \nu$ and $\rho \leftrightarrow \sigma$ whose expression is concretely given by
\begin{eqnarray}
&{}& [ \ddot g_{\rho\sigma}, \ddot g_{\mu\nu}^\prime ] = \kappa^2 [ w_1 ( \ddot g_{\rho\sigma} \dot g_{\mu\nu} 
- \ddot g_{\mu\nu} \dot g_{\rho\sigma}  )  + w_2 ( \delta_\rho^0 \delta_\sigma^0  \dddot{g}_{\mu\nu} 
- \delta_\mu^0 \delta_\nu^0  \dddot{g}_{\rho\sigma} )  ] \delta^3
\nonumber\\
&{}& + \kappa^2 w_3 ( \delta_\rho^0 \delta_\sigma^0 \ddot{g}_{\mu\nu} g^{0k} 
+ \delta_\mu^0 \delta_\nu^0  \ddot{g}_{\rho\sigma}^\prime g^{\prime 0k} ) \partial_k \delta^3
+ \kappa^2 \{ [ w_4 \dot g_{\rho\sigma} \dot g_{\mu\nu} + w_5 ( \dot g_{\rho\mu} \dot g_{\sigma\nu} 
+ \dot g_{\rho\nu} \dot g_{\sigma\mu}
\nonumber\\
&{}& + \dot g_{\sigma\mu} \dot g_{\rho\nu} + \dot g_{\sigma\nu} \dot g_{\rho\mu} )
+ w_6 ( \delta_\rho^0 \delta_\mu^0 \ddot g_{\sigma\nu} + \delta_\rho^0 \delta_\nu^0 \ddot g_{\sigma\mu} 
+ \delta_\sigma^0 \delta_\mu^0 \ddot g_{\rho\nu} + \delta_\sigma^0 \delta_\nu^0 \ddot g_{\rho\mu} )
\nonumber\\
&{}& + \delta_\rho^0 \delta_\sigma^0 \delta_\mu^0 \delta_\nu^0 ( w_7 \ddot g_{00} + w_8 (\dot g_{00})^2 ) ] g^{0k}
+ ( {\rm{unprimed}}  \rightarrow {\rm{primed}} ) \} \partial_k \delta^3.
\label{Form-ddot-g-ddot-g}
\end{eqnarray}
Inserting Eqs. (\ref{Nontrivial-sol}) and (\ref{Form-ddot-g-ddot-g}) to Eq. (\ref{Hi-Donder-ddot-g}), we can verify
that
\begin{eqnarray}
&{}& z ( x, x^\prime ) = 0,
\nonumber\\
&{}& [ \ddot g_{\rho\sigma}, \ddot g_{\mu\nu}^\prime ] = w (x, x^\prime) \delta_\rho^0 \delta_\sigma^0 \delta_\mu^0 \delta_\nu^0 \delta^3,
\label{Nontrivial-sol2}
\end{eqnarray}
where $w (x, x^\prime)$ is an appropriate function. Thus, from the former equation (\ref{Nontrivial-sol2}), we find that
\begin{eqnarray}
[ \ddot g_{\rho\sigma}, \dot g_{\mu\nu}^\prime ] = 0.
\label{ddot-g-d-g-ETC}
\end{eqnarray}
Repeating the similar procedure to the above, we can arrive at a proof of Eq. (\ref{zero-ETCR}).

\section{Vanishing commutation relation of metric tensor}

In our previous paper \cite{Oda-Can}, we have briefly commented on the physical meaning of Eq. (\ref{zero-ETCR}). In this section, 
we wish to elaborate on it from the perspective of a global $GL(4)$ symmetry and mention its implication for the renormalizability
of quadratic gravity. 

In order to understand what physical meaning the vanishing ETCRs (\ref{zero-ETCR}) have, let us consider a four-dimensional commutation 
relation between two metric tensors, and show that
\begin{eqnarray}
[ g_{\rho\sigma} (x), g_{\mu\nu} (x^\prime) ] = 0,
\label{4D-g-g-CR}  
\end{eqnarray}
for $x^\mu$ and $x^{\prime \mu}$ being spacelike separated.\footnote{Of course, there is a subtlety in this equation since one does not
know if $x^\mu$ and $x^{\prime \mu}$ are spacelike separated until one knows the metric tensor \cite{Wald}.} For this purpose, setting 
$x^\mu \equiv ( t, \vec{x} )$ and $x^{\prime \mu} \equiv ( t + \delta t, \vec{x}^\prime )$, and then expanding the LHS of Eq. (\ref{4D-g-g-CR}) 
around $x^0 = t$, we have
\begin{eqnarray}
[ g_{\rho\sigma} (x), g_{\mu\nu} (x^\prime) ] = [ g_{\rho\sigma}, g_{\mu\nu}^\prime]
+ \left[ g_{\rho\sigma}, \frac{\partial g_{\mu\nu}^\prime}{\partial t} \right] \delta t
+ \frac{1}{2} \left[ g_{\rho\sigma}, \frac{\partial^2 g_{\mu\nu}^\prime}{\partial t^2} \right] (\delta t)^2 + \cdots.
\label{4D-g-g-CR2}  
\end{eqnarray}
Thus, Eq. (\ref{zero-ETCR}) immediately leads to Eq. (\ref{4D-g-g-CR}). 

The key point is that Eq. (\ref{4D-g-g-CR}) is valid when $x^\mu$ and $x^{\prime \mu}$ are spacelike separated. This fact can be easily seen
from the space-time interval as
\begin{eqnarray}
(x^\prime - x)^2 &\equiv& g_{\mu\nu} (x^{\prime\mu} - x^\mu) (x^{\prime\nu} - x^\nu)
\nonumber\\
&=& g_{00} (\delta t)^2 + 2 g_{0i} \delta t (x^{\prime i} - x^i) + g_{ij} (x^{\prime i} - x^i) (x^{\prime j} - x^j)  
\nonumber\\
&\approx& g_{ij} (x^{\prime i} - x^i) (x^{\prime j} - x^j) > 0,  
\label{ST-inter}  
\end{eqnarray}
where we have assumed $| \delta t| \ll 1$. It is of interest that when $x^\mu$ and $x^{\prime \mu}$ are spacelike separated,
Eq. (\ref{4D-g-g-CR}) is nothing but the law of microcausality, i.e. a measurement of $ g_{\mu\nu}$ at $x^{\prime \mu}$ cannot influence
the value of $ g_{\mu\nu}$ at $x^\mu$ \cite{Wald}. 

Here it is important to recall that in the equal-time (anti)commutation relations (ETCRs), the ``time ${\it{t}}$'' is not the physical time, 
but generally a mere parameter which is needed to formulate the canonical formalism. In the formalism at hand, there is a global $GL(4)$ 
symmetry owing to the (higher-derivative) de Donder gauge condition. Then, the existence of the $GL(4)$ symmetry makes it possible to 
take an arbitrary linear combination of the space-time coordinates $x^\mu$ as such a time parameter. This fact implies that the meaning
of the space-time interval is so vague that we should regard the 4D CR in Eq. (\ref{4D-g-g-CR}) as being valid for arbitrary space-time
separation, i.e. spacelike, null and timelike separations. From this perspective, in quadratic gravity the metric tensor field 
behaves as if it were not a quantum operator but a classical field at least when we consider correlation functions 
between the metric tensors. 

On the other hand, in the other higher-derivative gravity such as conformal gravity \cite{Oda-Ohta, Oda-Conf} 
and $f(R)$ gravity \cite{Oda-f} as well as Einstein's general relativity \cite{Nakanishi, N-O-text}, Eq. (\ref{zero-ETCR}) does not hold. 
Thus, the vanishing ETCRs among the metric tensor and its derivatives are a peculiar feature of quadratic gravity, so it is tempting 
to conjecture that this phenomenon might be related to the renormalizability of quadratic gravity. 

Indeed, there is a folklore that we have an essential difference between quantum gravity and quantum field theory 
on a fixed background since in quantum gravity the metric field $g_{\mu\nu}$ plays the dual role as both the quantity which describes the 
dynamical aspects of gravity and the quantity which describes the background space-time structure \cite{Wald}. To put it differently,
in order to quantize the dynamical degrees of freedom associated with the gravitational field, we must also provide a quantum mechanical
description of space-time structure. This problem has no analog for the conventional quantum field theory formulated on a fixed, space-time 
background, which is treated classically. 

In the present formulation, as explained in Section 2, the dynamical information associated with the gravitational field is included in not
$\varphi_{\mu\nu}$ but $\phi, \psi_{\mu\nu}$ and $h_{\mu\nu}$.\footnote{As seen in Eq. (\ref{Background}),  $\varphi_{\mu\nu}$ 
corresponds to the metric fluctuation around the flat Minkowski metric $\eta_{\mu\nu}$.} This fact is consistent with the vanishing 
four-dimensional commutation relation of the metric, which implies that $\varphi_{\mu\nu}$ is not a quantum field but a classical background field. 
In other words, in the canonical operator formalism of quadratic gravity the metric tensor loses the dual role and consequently 
we do not encouter the difficulty of constructing a quantum field theory of gravity.

\section{Conclusions}

In this article, we have constructed the higher-derivative de Donder gauge in the manifestly covariant canonical 
operator formalism of quadratic gravity, particularly investigated the equal-time commutation relations (ETCRs) among 
the time derivatives of the metric tensor, and then found that they identically vanish even in the higher-derivative 
de Donder gauge as in the conventional de Donder gauge (or harmonic gauge) for general coordinate invariance. 
Though this result does not always mean that such the vanishing ETCRs are independent of a choice of the gauge fixing
condition, it supports for our expectation that this is in fact the case since we can at present derive all the ETCRs in 
closed form only by the de Donder gauge.

We have also verified that these ETCRs provide us with the vanishing four-dimensional commutation relation, which implies 
that the metric tensor behaves as if it were not a quantum operator but a classical field. In this case, the micro-causality 
is valid at least for the metric tensor in an obvious manner.  

Even if it is not described explicitly in this article, we have also attempted to investigate if some new ideas for the problem 
of the massive ghost, which are originally formulated in the path integral formalism or toy models, could be described 
in the framework of the canonical operator formalism, but in vain. Perhaps, they might be simply wrong or violate some fundamental postulates 
in quantum field theory such as causality or unitarity. Anyway, in the latter case we should clarify in future's work which postualtes 
are really violated or such a violation of postulates is physically allowed in the regime of quantum gravity or not.
 
It is a pity that we cannot yet find any new ideas which are consistent with the canonical operator formalism. 
However, since the canonical operator formalism gives us a sound foundation of quantum field theory, once we could find a solution
to the massive ghost problem, the solution would yield a correct solution to the problem in a direct manner. 
Thus, it might be valuable to pursue a new solution to the massive ghost problem within the framework of 
the canonical formalism. For instance, in the analysis of physical modes \cite{Oda-Can}, we usually assume that there is
no bound state. If this assumption is not imposed, we are free to take account of a bound state which is BRST-conjugate
to the massive ghost and they together consists of a BRST doublet, thereby belonging to unphysical state and nullifying
the massive ghost problem \cite{Kawasaki, Oda-Corfu}. We would like to return to the problem of the massive ghost
in near future.

\begin{flushleft}
{\bf Acknowledgements}
\end{flushleft}

We are grateful to D. Sorokin for valuable discussions. 
We would like to thank INFN Padova for financial support and warm hospitality.

\appendix
\addcontentsline{toc}{section}{Appendix~\ref{app:scripts}: Training Scripts}
\section*{Appendix}
\label{app:scripts}
\renewcommand{\theequation}{A.\arabic{equation}}
\setcounter{equation}{0}

\section{Alternative derivation of $[ \dot g_{\rho\sigma}, g_{\mu\nu}^\prime ] = 0$}

In this appendix, we will present an alternative derivation of Eq. (\ref{Ext-dot-g2}) in a direct manner without
using field equation.

Let us start with Eq. (\ref{Hi-Donder-g}), which can be rewritten into the form:
\begin{eqnarray}
[ I^\lambda{}_\rho, g_{\mu\nu}^\prime ] \partial_\sigma \tilde g^{\rho\sigma} 
+ I^\lambda{}_\rho [ \partial_\sigma \tilde g^{\rho\sigma}, g_{\mu\nu}^\prime ] = 0,
\label{App-Hi-Donder-g}
\end{eqnarray}
where $I^\lambda{}_\rho$ is defined in (\ref{I-lambda-rho}). In evaluating the commutator in the first term on the LHS 
of this equation, we use Eqs. (\ref{Ext-dot-g-g}), (\ref{Gamma-g}), (\ref{Ext-dot-g-dot-g3}) and (\ref{Riemann}) 
whose result reads
\begin{eqnarray}
[ I^\lambda{}_\rho, g_{\mu\nu}^\prime ] &=& \frac{1}{2} a x [ ( g^{\alpha\lambda} g^{0\beta} - g^{0\lambda} g^{\alpha\beta} ) 
\delta_\rho^0 \Gamma_{\alpha\beta}^0 + ( g^{0\lambda} g^{0\alpha} - g^{00} g^{\lambda\alpha} ) \Gamma_{\alpha\rho}^0 ] 
\delta_\mu^0 \delta_\nu^0 \delta^3   
\nonumber\\
&+& b x \delta_\rho^\lambda ( g^{0\alpha} g^{0\beta} - g^{00} g^{\alpha\beta} ) \Gamma_{\alpha\beta}^0 
\delta_\mu^0 \delta_\nu^0 \delta^3.   
\label{I-g-prime2}
\end{eqnarray}
Here note that terms dependent on $y (x, x^\prime)$ are cancelled out so they do not appear in the above expression.

The commutator in the second term on the LHS of Eq. (\ref{App-Hi-Donder-g}) has been already calculated in (\ref{de-Donder-g}).
Thus, Eq. (\ref{App-Hi-Donder-g}) can be calculated to 
\begin{eqnarray}
&{}& x \Bigg\{ \frac{1}{2} a [ ( g^{\alpha\lambda} g^{0\beta} - g^{0\lambda} g^{\alpha\beta} ) 
\delta_\rho^0 \Gamma_{\alpha\beta}^0 + ( g^{0\lambda} g^{0\alpha} - g^{00} g^{\alpha\lambda} ) \Gamma_{\alpha\rho}^0 ] 
\nonumber\\
&+& b \delta_\rho^\lambda ( g^{0\alpha} g^{0\beta} - g^{00} g^{\alpha\beta} ) \Gamma_{\alpha\beta}^0 \Bigg\}
 \partial_\sigma \tilde g^{\rho\sigma} \delta_\mu^0 \delta_\nu^0 \delta^3
- \frac{1}{2} x \tilde g^{00} g^{0\rho} ( a R^\lambda{}_\rho 
\nonumber\\
&+& b \delta_\rho^\lambda R ) 
\delta_\mu^0 \delta_\nu^0 \delta^3 = 0. 
\label{App-Hi-Donder-g2}
\end{eqnarray}
The only solution to this equation is $x = 0$, which means $[ \dot g_{\rho\sigma}, g_{\mu\nu}^\prime ] = 0$.



\begin{thebibliography}{99}

\bibitem{Stelle1}
K. S. Stelle, {``Renormalization of Higher Derivative Quantum Gravity'',
Phys. Rev. {\bf D 16} (1977) 953.}

\bibitem{Luca}
L. Buoninfante, {``Strict Renormalizability as a Paradigm for Fundamental Physics", 
arXiv:2504.05900 [hep-th] and references therein.}    

\bibitem{Anselmi}
D. Anselmi, {``On the Quantum Field Theory of the Gravitational Interactions", JHEP {\bf 06} 
(2017) 086.}    

\bibitem{Salvio}
A. Salvio, {``Quadratic Gravity'', Front. in Phys. {\bf 6} (2018) 77.}

\bibitem{Strumia}
A. Strumia, {``Interpretation of Quantum Mechanics with Indefinite Norm'', MDPI Physics {\bf 1} (2019) 17.}

\bibitem{Donoghue1}
J. E. Donoghue and G. Menezes, {``Unitarity, Stability and Loops of Unstable Ghosts'', Phys. Rev. {\bf D 100} 
(2019) 105006.}

\bibitem{Donoghue2}
J. E. Donoghue and G. Menezes, {``On Quadratic Gravity'', Nuovo. Cim. {\bf C 45} (2022) 26.}

\bibitem{Kubo-Kugo}
J. Kubo and T. Kugo, {"Unitarity Violation in Field Theories of Lee-Wick's Complex Ghost", PTEP {\bf 2023} 
(2023) 123B02.}

\bibitem{Holdom}
B. Holdom, {``Making Sense of Ghosts'', Nucl. Phys. {\bf B 1008} (2024) 116696.}

\bibitem{Luca2}
L. Buoninfante, {``Remarks on Ghost Resonances", 
JHEP {\bf 02} (2025) 186.}    

\bibitem{Kugo-Ojima}
T. Kugo and I. Ojima, {"Local Covariant Operator Formalism of Nonabelian Gauge Theories
and Quark Confinement Problem", Prog. Theor. Phys. Suppl. {\bf 66} (1979) 1.}

\bibitem{Nakanishi}
N. Nakanishi, {"Indefinite Metric Quantum Field Theory of General Gravity", 
Prog. Theor. Phys. {\bf 59} (1978) 972.}

\bibitem{N-O-text}
N. Nakanishi and I. Ojima, {"Covariant Operator Formalism of Gauge Theories and Quantum Gravity", 
World Scientific Publishing, 1990 and references therein.}

\bibitem{Kimura1}
S. Kawasaki, T. Kimura and K. Kitago, {"Canonical Quantum Theory of Gravitational Field with Higher Derivatives", 
Prog. Theor. Phys. {\bf 66} (1981) 2085.}

\bibitem{Kimura2}
S. Kawasaki and T. Kimura, {"Canonical Quantum Theory of Gravitational Field with Higher Derivatives. II", 
Prog. Theor. Phys. {\bf 68} (1982) 1749.}

\bibitem{Kimura3}
S. Kawasaki and T. Kimura, {"Canonical Quantum Theory of Gravitational Field with Higher Derivatives. III", 
Prog. Theor. Phys. {\bf 69} (1983) 1015.}

\bibitem{Oda-Can}
I. Oda, {``Manifestly Covariant Canonical Formalism of Quadratic Gravity", arXiv:2505.09149 [hep-th].}    

\bibitem{Oda-Q}
I. Oda, {``Quantum Scale Invariant Gravity in de Donder Gauge'', Phys. Rev. {\bf D 105} (2022) 066001.}   

\bibitem{Oda-W}
I. Oda, {``Quantum Theory of Weyl Invariant Scalar-tensor Gravity'', Phys. Rev. {\bf D 105} (2022) 120618.}   

\bibitem{Oda-Saake}
I. Oda and P. Saake, {"BRST Formalism of Weyl Conformal Gravity", 
Phys. Rev. {\bf D 106} (2022) 106007.}    

\bibitem{Oda-Corfu}
I. Oda, {``BRST formalism of Weyl Invariant Gravity and Confinement of Massive Tensor Ghost'', 
PoS CORFU2023 (2024) 158.}   

\bibitem{Oda-Ohta}
I. Oda and M. Ohta, {``Quantum Conformal Gravity", JHEP {\bf 02} (2024) 213.}    

\bibitem{Oda-Conf}
I. Oda, {``Conformal Symmetry in Quantum Gravity", Eur. Phys. J.  {\bf C 84} (2024) 887.}    

\bibitem{Oda-f}
I. Oda, {``BRST Formalism of $f(R)$ Gravity", arXiv:2410.20270 [hep-th].}    

\bibitem{Fradkin}
E. S. Fradkin and A. A. Tseytlin, {``Renormalizable Asymptotically Free Quantum Theory of Gravity'', 
Nucl. Phys. {\bf B 201} (1982) 469.}

\bibitem{NO-grav}
N. Nakanishi and I. Ojima, {``Proof of the Exact Masslessness of Gravitons'', 
Phys. Rev. Lett. {\bf 43} (1979) 91.}

\bibitem{MTW}
C. W. Misner, K. S. Thorne and J. A. Wheeler, {``Gravitation", W H Freeman and Co (Sd), 1973.}

\bibitem{Barth}
N. H. Barth, S. M. Christensen, {"Quantizing fourth order gravity theories. 1. The functional integral", 
Phys. Rev. {\bf D 28} (1983) 1876.}    

\bibitem{Ohta}
N. Ohta, {``General Procedure of Gauge Fixings and Ghosts'', Phys. Lett. {\bf B 811} (2020) 135965.}

\bibitem{Buchbinder}
I. L. Buchbinder and I. L. Shapiro, {``Introduction to Quantum Field Theory with Applications to
Quantum Gravity", Oxford University Press, 2021.}

\bibitem{Wald}
R. M. Wald, {``General Relativity'', The University of Chicago Press, 1984.}

\bibitem{Kawasaki}
S. Kawasaki and K. Kimura, {"A Possible Mechanism of Ghost Confinement in a Renormalizable 
Quantum Gravity", Prog. Theor. Phys. {\bf 65} (1981) 1767.}






\end{thebibliography}
\end{document}